\documentclass[reprint,amsmath,amssymb,aps,nofootinbib]{revtex4-1}

\usepackage{graphicx}
\usepackage{dcolumn}
\usepackage{bm}
\usepackage{natbib}
\usepackage{empheq}

\newcommand{\p}{\partial}
\newcommand{\half}{\frac{1}{2}}

\newcommand\nn{{\nonumber}}
\newcommand{\beq}{\begin{equation}}
\newcommand{\eq}{\end{equation}}

\def\bea{\begin{eqnarray}}
\def\ea{\end{eqnarray}}
\def\p{\partial}

\begin{document}

\preprint{DAMTP-2013-58}

\title{Spatially Modulated Instabilities for Scaling Solutions at Finite Charge Density}
 
\author{Sera Cremonini$ ^{\,\clubsuit,\spadesuit}$}
\affiliation{\it $ ^\clubsuit$
\it DAMTP, Centre for Mathematical Sciences, \\ University of Cambridge, Wilberforce Road, Cambridge, CB3 0WA, UK \\
\it $ ^\spadesuit$ George and Cynthia Mitchell Institute for Fundamental Physics and Astronomy,\\
\it Texas A\&M University, College Station, TX 77843--4242, USA}

\date{\today}

\begin{abstract}
We consider finite charge density geometries which interpolate between $AdS_2 \times \mathbb{R}^2$ 
in the infrared and $AdS_4$ in the ultraviolet, while traversing an intermediate regime of 
anisotropic Lifshitz scaling and hyperscaling violation.
We work with Einstein-Maxwell-dilaton models and only turn on a background electric field.
The spatially modulated instabilities of the near-horizon $AdS_2$ part of the geometry are used to 
argue that the scaling solutions themselves should be thought of as being unstable -- in the deep infrared -- 
to spatially modulated phases. 
We identify instability windows for the scaling exponents $z$ and $\theta$, 
which are refined further by requiring the solutions to satisfy the null energy condition.
This analysis reinforces the idea that, for large classes of models, spatially modulated phases describe 
the ground state of hyperscaling violating scaling geometries.
\end{abstract}

\pacs{Valid PACS appear here}

\maketitle

\section{Introduction}
\label{Intro}

Recent years have seen growing interest in applying AdS/CFT methods to condensed matter systems 
whose underlying degrees of freedom are strongly coupled -- notoriously  
difficult to explore using traditional methods.
New gravitational solutions and instabilities have provided a rich playground for 
describing novel phases of quantum matter whose behavior is poorly understood.
As an example, classes of scaling geometries which break Lorentz invariance have been used to model
some of the unconventional properties observed in non-Fermi liquids and
strongly correlated electron systems.

Lately the focus has shifted to probing and classifying 
gravitational solutions with spatial anisotropy 
and/or inhomogeneities,
motivated by new  qualitative and quantitative insights into transport in systems with broken translational invariance.
A rich structure has emerged from holographic realizations of spatially modulated phases 
(see \cite{Domokos:2007kt,Nakamura:2009tf,Flauger:2010tv,Donos:2011bh} for early work), 
which appear in condensed matter in a number of settings -- 
nematics, smectics and charge/spin density waves being prime examples.
Interestingly, some of the anisotropic ground states resulting from the breaking 
of translational invariance 
have played a key role in recent attempts to model holographically the formation of a crystal structure 
(see \emph{e.g.} \cite{Bao:2013fda}).
We refer the reader to \cite{Rozali:2012es,Donos:2013wia,Withers:2013loa,Withers:2013kva,Rozali:2013ama} 
for recent constructions of the inhomogeneous geometries resulting from the backreaction 
of spatially modulated perturbations.

In this note we are interested in the deep infrared fate of a class of (four-dimensional) 
solutions which exhibit anisotropic Lifshitz scaling and hyperscaling violation. 
As we will see shortly, they arise as exact solutions to simple Einstein-Maxwell-dilaton models 
\cite{Gubser:2009qt,Cadoni:2009xm,Charmousis:2010zz,Perlmutter:2010qu,Iizuka:2011hg,Gouteraux:2011ce,Huijse:2011ef}.
More generally, however, they should be thought of as arising in the intermediate, `mid-infrared' region of more complicated 
geometries, which typically flow in the infrared to either 
$AdS_2 \times \mathbb{R}^2$ 
(as emphasized in \cite{Harrison:2012vy,Bhattacharya:2012zu,Kundu:2012jn,Knodel:2013fua}) 
or a spacetime conformal to it. 
In the former case, the extensive entropy associated with the $AdS_2 \times \mathbb{R}^2$ near-horizon 
description raises the question of what is the nature of their true ground state.

It is by now well-known that $AdS_2 \times \mathbb{R}^2$ suffers from spatially 
modulated instabilities in a number of constructions (see e.g. \cite{Donos:2011pn,Donos:2011qt}).
The presence of such unstable modes suggests that, in appropriate regions of phase space, 
the endpoint of scaling solutions with a `naive' $AdS_2$ infrared completion 
should also be spatially modulated phases.
This logic was advocated in \cite{Cremonini:2012ir} for anisotropic, hyperscaling violating
solutions supported by a constant magnetic field.
It was also corroborated by the complementary analysis of \cite{Iizuka:2013ag}, which identified striped instabilities by examining
the scaling geometries directly (without assuming a flow to $AdS_2$ in the IR), 
but relied crucially on the presence of an axionic term.
In fact in all of these cases the presence of instabilities was directly tied to terms that violated 
time reversal (T) and parity (P) invariance.
However, it emphasized recently in \cite{Donos:2013gda} that T and/or P violation are in fact \emph{not} needed 
in order for $AdS_2 \times \mathbb{R}^2$ to become unstable to spatially modulated perturbations.

Here we revisit the analysis of \cite{Cremonini:2012ir} and apply it to scaling solutions at finite charge density, 
more relevant to the condensed matter context and in particular to compressible phases of matter.
As in \cite{Cremonini:2012ir}, we will rely on the assumption that these scaling solutions are approaching 
$AdS_2 \times \mathbb{R}^2$ in the infrared.
Building on \cite{Donos:2013gda}, we will identify the onset of 
spatially modulated instabilities for certain classes of anisotropic, hyperscaling violating solutions.
Ensuring that null energy conditions are satisfied will refine the analysis further.
As we will see, for large classes of models phases with stripe order arise quite generically 
as the natural infrared description of these scaling solutions.

\section{The Setup}
\label{Setup}

We work with four-dimensional Einstein-Maxwell-dilaton (EMD) gravity
\beq
\label{EMDLag}
\mathcal{L} = R  - \half (\p\phi)^2 - f(\phi) F_{\mu\nu}F^{\mu\nu} - V(\phi)\, ,
\eq
with the scalar potential $V(\phi)$ and the gauge kinetic coupling $f(\phi)$ for now left entirely arbitrary.
The equations of motion for this system are
\bea
&R_{\mu\nu}& =  \half \left(\p_\mu \phi \, \p_\nu \phi + V g_{\mu\nu} \right) - \half f 
\left( g_{\mu\nu} F^2 +  4  F_{\mu\rho} F^{\rho}_{\;\;\nu} \right),  \nn \\
& \nabla_\mu & \left( f F^{\mu\nu} \right) = 0 \, , \qquad \Box \phi - V^\prime - f^\prime F^2 = 0 .
\ea
The background gauge field is taken to be purely electric, with $A_t = Q_e A(r)$ and all other components vanishing.
We are interested in zero temperature geometries which interpolate between $AdS_2 \times \mathbb{R}^2$
in the deep IR and $AdS_4$ in the UV, while traversing \emph{an intermediate scaling region} described by
\beq
\label{scalingmetric}
ds^2 = r^{-2+\theta} \left(-r^{-2(z-1)} dt^2 + dr^2 + d\vec{x}^2 \right) \, .
\eq
In addition to the dynamical critical exponent $z$, the metric (\ref{scalingmetric}) 
is characterized by a hyperscaling  violating exponent $\theta$, thanks to which 
it is no longer scale invariant 
but transforms as $ds^2 \rightarrow \lambda^\theta ds^2$ under $t \rightarrow \lambda^z t$, $r \rightarrow \lambda \, r$,
$x_i \rightarrow \lambda \, x_i$. 
The intermediate geometry is supported by a running scalar, $\phi \sim \log r$, which breaks the exact Lifshitz symmetry of the metric
-- hence it is only `Lifshitz-like.'

The exponents $\{z,\theta\}$ modify the `usual' scalings of thermodynamic quantities, giving \emph{e.g.}
$s \sim T^{(d-\theta)/z}$ for the entropy of a $(d+1)$-dimensional field theory. 
The case $d-\theta=1$ has received particular attention because it leads to 
logarithmic violations of the area law of entanglement entropy, $S_{ent} \sim A \, log A$, 
a tell-tale of the presence of a Fermi surface \cite{Huijse:2011ef,Ogawa:2011bz}.
More generally, $z$ and $\theta$ should be thought of as tunable parameters, and (\ref{scalingmetric}) 
as a useful laboratory to reproduce particular scalings of systems of interest. 
 
The scaling geometries (\ref{scalingmetric})  arise as \emph{exact solutions} to the model (\ref{EMDLag}) when 
\beq
f \sim e^{\alpha \phi} \qquad \text{and} \qquad V \sim e^{-\eta\phi} \, ,
\eq
with $\{z,\theta\}$ related to the Lagrangian parameters 
$\{\alpha,\beta\}$ through 
\beq
\label{ztheta}
\theta = \frac{4\eta}{\alpha+\eta} \, , \qquad z = 1 + \frac{\theta}{2} + \frac{(4-\theta)^2}{2(2-\theta)\, \alpha^2} \, .
\eq
In turn these can be inverted to give
\beq
\alpha^2 = \frac{(\theta-4)^2}{(\theta-2)(\theta-2z+2)} \, , \qquad \eta = \frac{\theta \alpha}{4-\theta} \, .
\eq
Finally, imposing the null energy conditions (NEC) \cite{Dong:2012se} in the intermediate region 
further constrains the physically allowed values of $z$ and $\theta$,
\bea
\label{NEC}
&& (\theta-2)(2-2z+\theta) \geq 0 \, , \nn \\
&& (z-1)(2+z-\theta)  \geq 0 \, .
\ea 

{\bf The deep infared.}
In the IR we require the scalar to settle to a constant, $\phi=\phi_0$, 
and the metric to become that of $AdS_2 \times \mathbb{R}^2$,
\beq
ds^2 = L^2 \left( -r^2 dt^2 + \frac{dr^2}{r^2} + d\vec{x}^2 \right) \, .
\eq
The equations of motion then yield the following near-horizon 
conditions on the gauge kinetic function and scalar potential,
\bea
\label{cond1}
&& V(\phi_0) = - \frac{1}{L^2} \, , \\
\label{cond2}
&& V^\prime (\phi_0) f(\phi_0) = - V(\phi_0) f^\prime(\phi_0)  \, , \\
\label{cond3}
&& Q_e^2 = - \frac{1}{2 V(\phi_0) f(\phi_0)} \, .
\ea
We see from (\ref{cond1}) and (\ref{cond3}) that $V(\phi_0)<0$ and $f(\phi_0)>0$
are needed to ensure that the charge $Q_e$ and the AdS radius $L$ are real.

{\bf The ultraviolet.}
In order to operate within the standard holographic framework, in the deep ultraviolet we are interested in 
solutions which approach $AdS_4$ with a constant scalar $\phi=\phi_{UV}$.
The UV value of the scalar is then determined entirely by the condition that the effective scalar potential
$V_{eff} = V(\phi) + f(\phi) \, F^2$ admits an extremum, $\p_\phi V_{eff} (\phi_{UV}) = 0$.
However, we should emphasize that the structure of the instabilities and the main point of this analysis are 
linked to the infrared portion of the geometry and are largely insensitive to the UV.
Thus, the main results of this note will not be directly affected by the choice of other UV fixed points.

\section{Spatially Modulated Instabilities}
\label{SectionInstabilities}

The spatially modulated instabilities of electrically charged $AdS_2 \times \mathbb{R}^2$
solutions to the class of models (\ref{EMDLag}) were analyzed recently in \cite{Donos:2013gda}.
Here we build on the final results of their analysis and identify instability windows for 
the parameters of the theory -- first by 
working in a small-momentum approximation and then by considering a few simple exact cases.
In Section \ref{SectionIntermediate} we will apply this instability analysis
to $\{z,\theta\}$ scaling solutions assumed to have an $AdS_2 \times \mathbb{R}^2$ infrared completion.

After turning on the following set of time dependent, spatially modulated linear fluctuations
of the $AdS_2 \times \mathbb{R}^2$ background geometry,
\bea
&& \delta \phi = e^{-i\omega t} h (r) \cos k x_1 \, , \nn \\
&& \delta g_{tt} = L^2 r^2 e^{-i\omega t} h_{tt}(r) \cos k x_1 \, , \nn \\
&& \delta g_{x_i x_i} = L^2 e^{-i\omega t} h_{x_i x_i}(r) \cos k x_1 \, , \nn \\
&& \delta g_{t x_1} = L^2  e^{-i\omega t} h_{t x_1}(r) \sin k x_1 \, , \nn \\
&& \delta A_t = - E e^{-i\omega t} a_t(r) \cos k x_1 \, , \nn \\
&& \delta A_{x_1} = - E e^{-i\omega t} a_{x_1}(r) \sin k x_1 \, ,
\ea
with $i=\{1,2\}$, and using the remaining gauge freedom to identify \emph{three} gauge invariant combinations
$\vec{v} = \{\Phi_1,\Phi_2,\Phi_3\}$, the fluctuation equations take the form \cite{Donos:2013gda}
of three mixed modes propagating on $AdS_2$,
\beq
\left(  \frac{\omega^2}{r^2} + r^2 \p_r^2 + 2 r \p_r \right) \vec{v} = M^2 \vec{v} \, .
\eq
The mass matrix $M$ is given by\footnote{We are using the mass matrix notation of \cite{Donos:2013gda}.} 
\beq
M^2 = \begin{bmatrix}
   {2+ 2 \tau_1^2 +k^2} & {\; \; -2k^2 \;\;} & {2 \tau_1 (2-k^2 - \tau_2 -v_2)}  \\
  {-1} & {k^2} & {-2 \tau_1} &  \\
   { - \tau_1 } & 0 & {k^2 +v_2 + \tau_2}
    \end{bmatrix}
\eq
where the parameters $\tau_1,\tau_2, v_2$ are defined \cite{Donos:2013gda} through the expansions of the gauge kinetic function and 
scalar potential around the infrared value of the scalar $\phi_0$,
\bea
\label{expansionf}
f &=& f_0 \left( 1+ \tau_1 (\phi-\phi_0) -\frac{\tau_2}{2}(\phi-\phi_0)^2 + \ldots \right)  , \\
V &=& v_0 \left( 1- \tau_1 (\phi-\phi_0) -\frac{v_2}{2}(\phi-\phi_0)^2 + \ldots \right)  . 
\label{expansionV}
\ea
Notice that $f_0 >0$ and $v_0 <0$, and the equations of motion (\ref{cond1})--(\ref{cond2}) were used to relate to each other the terms linear in $\phi$.

Spatially modulated instabilities in this system are present when, at finite momentum, at least one of the mass
eigenvalues violates the $AdS_2$ BF bound, i.e. 
when $m_i^2 < -\frac{1}{4}$.
From the structure of the mass matrix we immediately see that the
eigenvalues $m^2_i$ are controlled by $\tau_1$ and the combination $(\tau_2 + v_2)$.
Finally, the fact that these instabilities are triggered without the need for $P$ or $T$ violation 
is reflected by the structure of the mass matrix, which depends only on $k^2$
and not on $k$. It is also reflected by the fact that the `dangerous' modes are 
\emph{static} and correspond to $\delta g_{t x_1} = \delta A_{x_1} =0$ \cite{Donos:2013gda}.

\subsection{Conditions for Instabilities}

In the zero momentum case the mass matrix simplifies and one finds
\beq
\label{zerokeigen}
m^2_i = \{0,\; 2, \; 2\tau_1^2 + \tau_2 + v_2 \} \, .
\eq
Thus, the system will have unstable modes already at $k=0$ when the parameters are such that
\beq
\label{zerokinst}
2 \tau_1^2 + \tau_2 + v_2 < - \frac{1}{4} \, .
\eq
However, these perturbations do not break translation invariance and should indicate instabilities 
to the formation of other $AdS_2 \times \mathbb{R}^2$ solutions or geometries conformal to it.
Since here we are interested in (stripe) instabilities triggered at finite momentum,
we will work under the assumption that (\ref{zerokinst}) is never satisfied, so that all the eigenvalues
(\ref{zerokeigen}) are guaranteed to be above the $AdS_2$ BF bound.

At finite momentum the eigenvalues can be solved for exactly, but are significantly cumbersome.
We start by working in a small momentum approximation,
which will be enough to illustrate the main point we are after.
For a couple of special parameter choices, we will also make use of the exact eigenvalues.

{\bf Small Momentum Expansion.}
Assuming that the matrix eigenvalues have a momentum expansion of the form
$$\lambda = \lambda_0 + k^2 \lambda_1 + {\cal O}(k^4) \, , $$
and using the fact that we know their zero-momentum values $\lambda_0$ from (\ref{zerokeigen}),
we find (provided that $\tau_2 + v_2 \neq -2\tau_1^2$) the following expressions:
\bea
\label{finitekeigen}
&& m_1^2 = 0 + {\cal O}(k^4)\, , \\
&& m_2^2 = 2 + k^2 \left(\frac{2\tau_1^2 + 2 (\tau_2 + v_2) -4 }{2\tau_1^2 + (\tau_2 + v_2) -2}\right) +{\cal O}(k^4) \, , \\
&& m_3^2 = 2 \tau_1^2 + (\tau_2 + v_2) + k^2 \left(\frac{4 \tau_1^2 +
(\tau_2 + v_2) -2 }{2\tau_1^2 + (\tau_2 + v_2) -2 }  \right) \nn \\ && + {\cal O}(k^4) \, .
\ea
We will come back to the case $\tau_2 + v_2 = -2\tau_1^2$, which needs to be analyzed separately, shortly.
Notice that to see potential instabilities associated with the first eigenvalue becoming negative
we must go to higher order in momentum,
\bea
m_1^2 &=& \half \left(\frac{\tau_2 + v_2}{2\tau_1^2+\tau_2 + v_2}\right) k^4 + \nn \\
&& \left(\frac{2\tau_1^2 - (\tau_2 + v_2) \tau_1^2 -(\tau_2 + v_2)^2}{2(2\tau_1^2+\tau_2 + v_2)^2}\right) \, k^6
+ \ldots \nn
\ea
Here we will content ourselves with examining the structure of the remaining two eigenvalues, 
neglecting terms of order ${\cal O}(k^4)$ and higher. We don't expect any qualitative differences by including higher order
terms in momentum.

We will take the onset of the instability to be signaled by the condition that the $k^2$ dependent terms become
negative -- the logic being that for an appropriate value of the momentum $k=k_\ast$, this contribution will win over the
leading $k=0$ term, pushing at least one of the mass eigenvalues below the $AdS_2$ BF bound\footnote{Clearly
this has to be done within the regime of validity of the small $k$ eigenvalue approximation we are employing.}.
In particular, examining both $m_2$ and $m_3$ we see that the $k$-dependent terms become negative when
\beq
\label{window}
\boxed{ \;
1- \half (\tau_2 + v_2) < \tau_1^2 < 2- (\tau_2 + v_2) \; }
\eq
which we therefore identify with the `window' for instabilities (provided that $\tau_2 + v_2 \neq -2\tau_1^2$).
Note that from this relation we learn that (small $k$) unstable modes are only possible for $ \tau_2 + v_2 <2$.

{\bf Exact Eigenvalues.}
For certain parameter choices the mass eigenvalues are easy to analyze exactly, without resorting to any small momentum expansion:
\begin{itemize}
\item
\emph{Case 1.} The case $\tau_2 + v_2 = -2\tau_1^2$, which was omitted from the small $k$ analysis above, gives
\beq
m_i^2 = \left\{ k^2 \, , 1 + k^2  \pm \sqrt{1 + 2k^2 (1+\tau_1^2)}\; \; \right\} \, .
\eq
The last eigenvalue is the only one which can dip below the $AdS_2$ BF bound, in fact it attains its lowest
value at $k^2_{\ast} = \frac{\tau_1^2 (2+\tau_1^2)}{2(1+\tau_1^2)}$, where it equals
\beq
m^2_{min} = -\half \frac{\tau_1^4}{1+\tau_1^2} \, .
\eq
Violations of the $AdS_2$ BF bound and therefore instabilities will occur whenever $ m^2_{min} < - \frac{1}{4}$,
which translates to 
\beq
\label{Ex1inst}
\boxed{\tau_1^2 >1 \, . \;}
\eq

\item
\emph{Case 2.}
Another special parameter choice is $\tau_2 + v_2 =2$, which as we noted above
`closes' the instability window (\ref{window}).
In this case the mass matrix simplifies significantly and we have
\bea
m_1^2 &=& k^2 +2 \, , \nn \\
m_{2,3}^2 &=& 1 + k^2  + \tau_1^2 + \pm \sqrt{1 + 2k^2 (1+\tau_1^2)+ 2\tau_1^2 + \tau_1^4} \nn 
\ea
All the squared-mass eigenvalues are now non-negative (the third one attains its minimum at $k=0$) and therefore we don't
see any unstable modes, as anticipated from (\ref{window}).
\end{itemize}
A more exhaustive analysis of the exact mass eigenvalues is beyond the scope of this paper.

\section{The intermediate scaling regime}
\label{SectionIntermediate}

So far we have restricted our attention to the instabilities of 
$AdS_2 \times \mathbb{R}^2$ solutions to the class of models (\ref{EMDLag}).
However, what we are after is what they can teach us about the true ground state of the 
scaling solutions (\ref{scalingmetric}).
Recall that our interest is in geometries which contain an intermediate 
$\{z, \theta\}$ scaling branch, and relax to $AdS_2 \times \mathbb{R}^2$ only in the deep IR. 
For solutions of this type, we can use the instability analysis of Section \ref{SectionInstabilities} to 
argue that the scaling geometries themselves should be thought of as being unstable -- in the deep infrared -- to spatially 
modulated phases. 
The relations (\ref{window}) and (\ref{Ex1inst}) then translate into conditions on the values of $z$ and $\theta$ associated with 
infrared instabilities.

In order to introduce the `minimal' set of ingredients needed to support
the geometries (\ref{scalingmetric}), we assume a simple gauge kinetic function of the form
\beq
f(\phi) = e^{\alpha \phi} \, .
\eq
With this choice, the coefficients $\tau_1$ and $\tau_2$ 
defined in (\ref{expansionf}) are just
\beq
\tau_1 = \alpha \, , \qquad \tau_2 = -\alpha^2 \, .
\eq
Next, we take the scalar potential to be of the form
\beq
V (\phi)= V_0 \, e^{-\eta\phi} + \mathcal{V} (\phi) \, ,
\eq
where we assume that the first term is responsible for driving the intermediate hyperscaling violating regime 
while $\mathcal{V} (\phi)$ is  negligible there. Given these assumptions, 
the exponents $z$ and $\theta$ can then be gotten from $\alpha$ 
and $\eta$ by using (\ref{ztheta}).
Finally, expanding the potential about $\phi_0$ we extract
\beq
\label{v2inst}
v_2 = L^2 \left(\eta^2 V_0 e^{-\eta\phi_0} + \mathcal{V}^{\prime\prime} \right)
= L^2 \left(\mathcal{V}^{\prime\prime}  - \eta^2 \mathcal{V} \right) -\eta^2 ,
\eq
where we made use of (\ref{cond1}) and it is understood that $\mathcal{V}$ and 
$\mathcal{V}^{\prime\prime} $ are evaluated at $\phi_0$.

\begin{figure}[t]
\begin{center}
\includegraphics*[width=3.0 in]{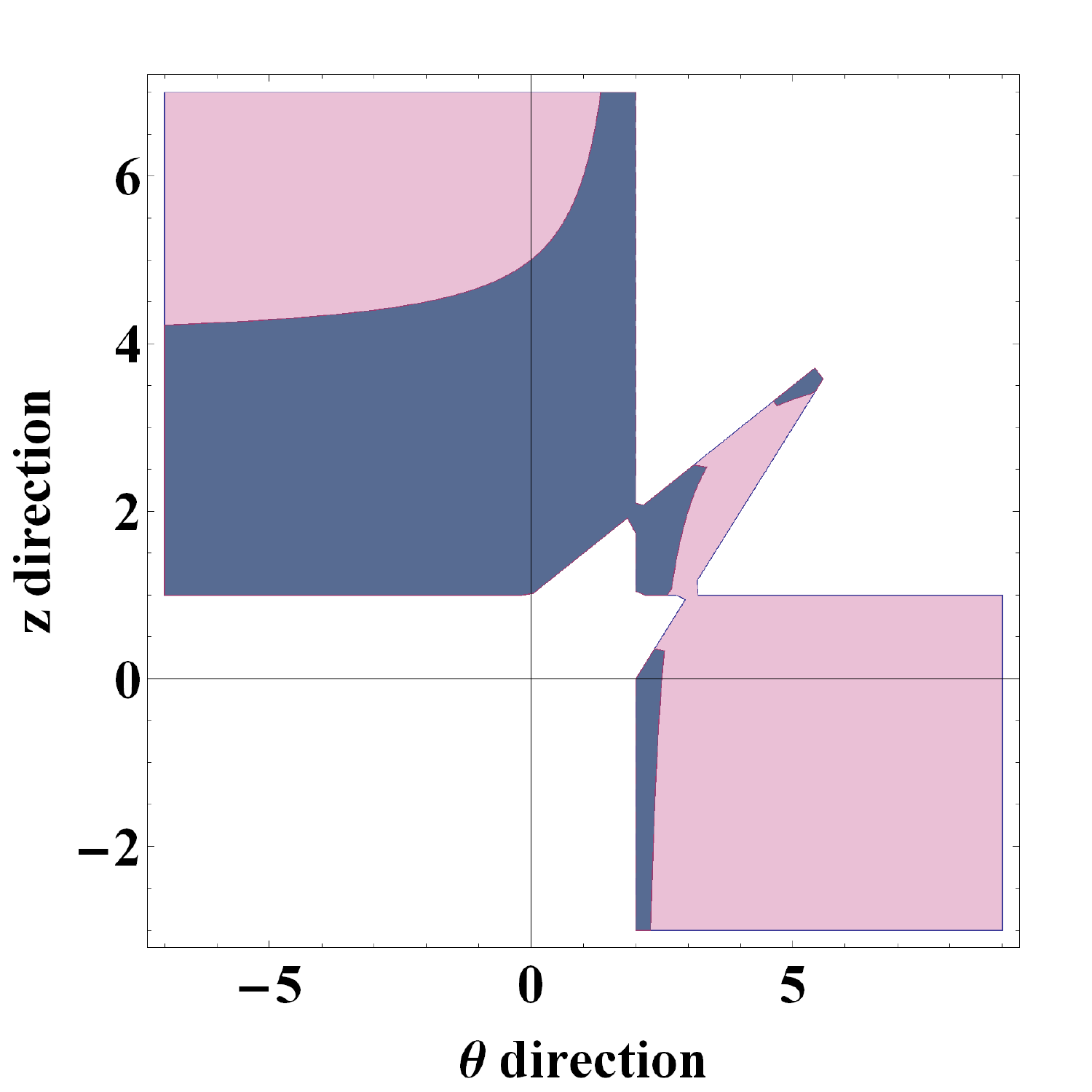} \label{Figure1}
\hspace{0.2cm}
\end{center}
\caption{The dark (blue) shaded region represents the values of $z$ and $\theta$ 
compatible with NEC and associated with the spatially modulated instabilities predicted by 
(\ref{case1}) for \emph{Case 1}. 
The lighter (pink) shaded regions contains the remaining values of $\{z,\theta\}$ allowed by NEC but outside 
the instability window (\ref{case1}).
We are plotting the range $-7<\theta<9$, $-3<z<7$.}
\end{figure}

Looking back at the small-momentum instability window (\ref{window}) we derived in Section \ref{SectionInstabilities}, 
it can now be rewritten as
\beq
\alpha^2 > 2- v_2 > 0,
\eq
or, entirely in terms of $z$, $\theta$ and $\mathcal{V}$:
\begin{empheq}[box=\fbox]{align}
& \;\; \frac{(\theta-4)^2}{(\theta-2)(\theta-2 z +2)} > 2 -L^2 \mathcal{V}^{\prime\prime}(\phi_0) + \nn \\
& \; \; \;\; + \; \frac{\theta^2}{(\theta-2)(\theta-2 z +2)} \, \left[ 1+L^2 \mathcal{V}(\phi_0) \right] > 0 .\;\;
\label{approx}
\end{empheq}
Scaling geometries with an infrared $AdS_2$ completion and whose $\{z,\theta\}$ exponents satisfy this inequality -- for appropriate values of 
$\mathcal{V}(\phi_0) $ and $\mathcal{V}^{\prime\prime}(\phi_0)$-- will then be unstable to the formation of spatially 
modulated phases. The NEC conditions (\ref{NEC}) should also be imposed and 
further constrain
the range of $z$ and $\theta$, as we show below.
Although we won't do it here, imposing thermodynamical stability would lead to additional restrictions.

Perhaps more interesting is \emph{Case 1}, which had a simple exact solution and exhibited instabilities when 
(\ref{Ex1inst}) was satisfied.
What this condition tells us is that scaling solutions with 
$z$ and $\theta$ 
in the range
\beq
\label{case1}
\boxed{\; \frac{(\theta-4)^2}{(\theta-2)(\theta-2 z +2)} >1 \;}
\eq
will have spatially modulated fluctuations provided $\mathcal{V}(\phi)$ satisfies the following relation:
\beq
\label{case1cond}
\frac{8}{\theta-2z + 2} = 
L^2 \left( \mathcal{V}^{\prime\prime}(\phi_0)- \frac{\theta^2}{(\theta-2)(\theta-2z + 2)} \, \mathcal{V}(\phi_0) \right) .
\eq
We illustrate \emph{Case 1} in Fig. 1. The dark (blue) region contains the values of the exponents $z$ and $\theta$
compatible with (\ref{case1}) and with the null energy conditions (\ref{NEC}). Thus, it indicates the portion of phase space 
susceptible to instabilities for the analysis of \emph{Case 1}. 
The remaining light (pink) region represents values of $\{z,\theta\}$ allowed by NEC
but outside of this particular instability window (see \cite{Bueno:2012vx}
for a plot of NEC for models with more complicated matter content). 
Note that for this case, the `special' value $\theta =1$ 
-- associated with a log violation of the entanglement entropy in this number of dimensions --
 is associated with an unstable geometry for $1.5 \leq z \leq 6$.

In summary,
for appropriate scalar potential profiles there will be regions of phase space in which 
conditions such as (\ref{approx}), (\ref{case1}) and (\ref{case1cond}) are easily satisfied. 
This suggests that -- for the corresponding parameter choices -- 
phases with spatial modulation are indeed the ultimate ground states of these
classes of scaling solutions.

\subsection{Examples}

For concreteness, we consider a few explicit examples:
\begin{enumerate}
\item
When $\mathcal{V}^{\prime\prime}(\phi_0)  = \eta^2 \mathcal{V}(\phi_0)$, 
the expression (\ref{v2inst}) reduces to $v_2 = -\eta^2 $.
All the instabilities of the system then are controlled entirely by the values of $\alpha$ and $\eta$. 
This condition is clearly satisfied, 
for instance, when the full scalar potential is $V \sim \cosh\eta\phi$, a choice used in several 
constructions in the literature.

The small-$k$ instability window (\ref{approx}) then takes the simple form
\beq
\alpha^2 -\eta ^2 > 2 \, .
\eq
Expressing it in terms of $z$ and $\theta$, we see that we should expect spatially modulated instabilites triggered 
by small momentum modes when
\beq
\label{coshcase}
-4 < \theta - 2 z + 2 < 0 .
\eq
The NEC conditions further restrict the allowed range of $z$ and $\theta$, as shown in Fig. 2.
As previously, the dark (blue) shaded region contains the values of $z$ and $\theta$ compatible with NEC and susceptible 
to instabilities according to (\ref{coshcase}).
The remaining lighter (pink) region represents the additional $\{z,\theta\}$ values allowed by NEC but 
which fall outside of this particular instability window.

\begin{figure}[t]
\begin{center}
\includegraphics*[width=2.8 in]{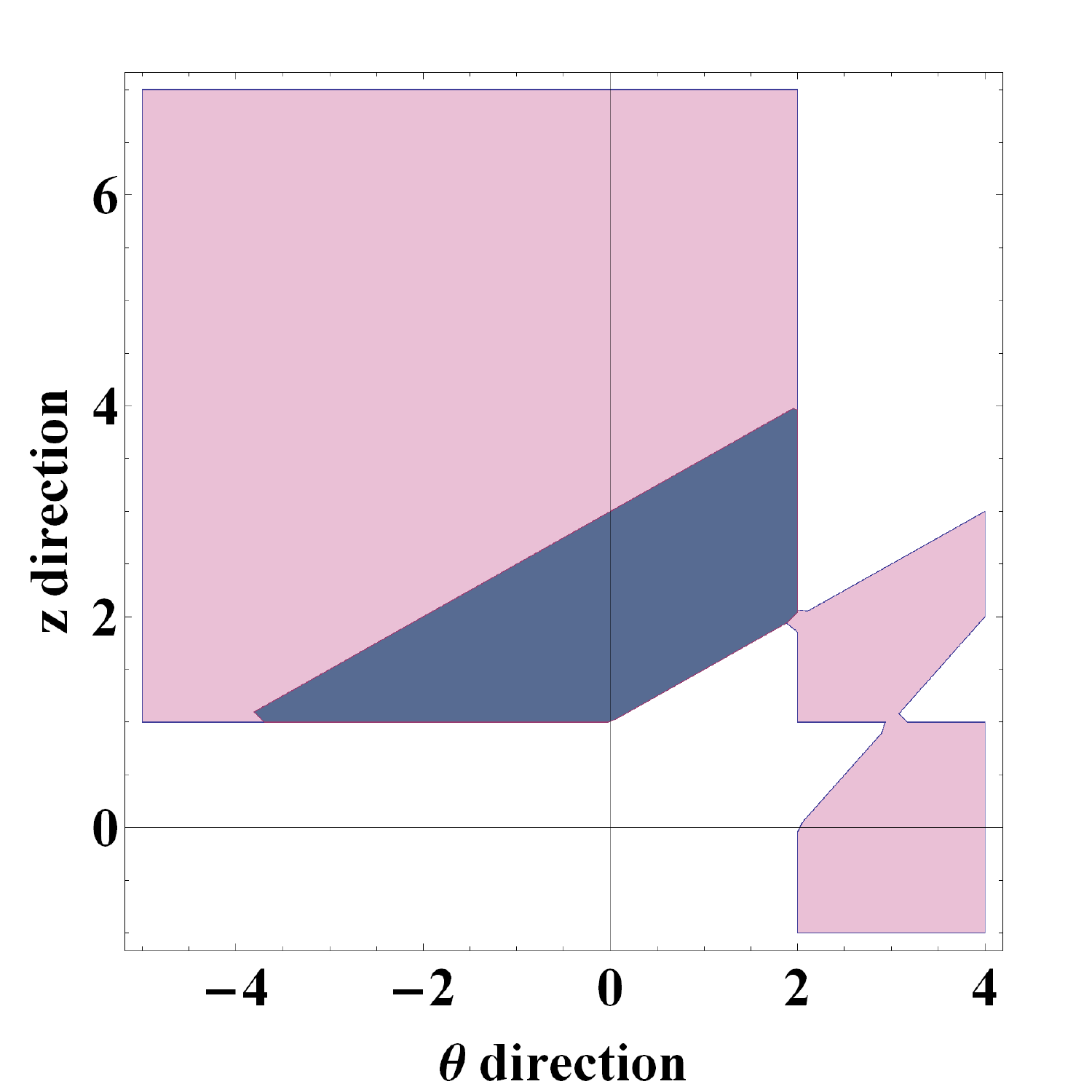}\label{Figure2}
\hspace{0.2cm}
\end{center}
\caption{The dark (blue) shaded region denotes the values of $z$ and $\theta$ compatible with NEC and susceptible 
to instabilities, for the special example described by (\ref{coshcase}). The light (pink) region 
contains the remaining values of $\{z,\theta\}$ allowed by NEC. }
\end{figure}

In the Lorentz invariant case $z=1$, the instability condition (\ref{coshcase}) becomes
\beq
- 4 <\theta < 0 \, ,
\eq
while when the hyperscaling violating exponent vanishes $\theta=0$ 
it becomes
\beq
1 < z < 3 \, .
\eq
Both conditions are automatically consistent with NEC.
Finally, for the special $\theta=1$ case the instability window becomes
\beq
\frac{3}{2} < z < \frac{7}{2} \, ,
\eq
also compatible with NEC.
Notice that all of these relations are clearly visible in Fig. 2.

\item
For a racetrack-type potential,
\beq
V(\phi) = V_0 e^{-\eta\phi} + V_1 e^{\gamma\phi} \,  ,
\eq
requiring as usual an IR $AdS_2 \times \mathbb{R}^2$ and after some manipulations we find $v_2 = \alpha (\gamma - \eta) - \eta \gamma$.
Small-$k$ perturbations then trigger instabilities when 
\beq
0 < 2 + \alpha (\eta -\gamma) + \eta\gamma < \alpha^2 \, ,
\eq
or equivalently when
\beq
0 < 2 + \frac{\theta \left[ 4(1-\kappa) +\theta(2\kappa-1)\right]}{(\theta-2)(\theta-2z+2)} < 
\frac{(\theta-4)^2}{(\theta-2)(\theta-2z+2)},
\eq
where we defined $\kappa \equiv \gamma/\eta$ for convenience. 

On the other hand, the exact \emph{Case 1} of Section \ref{SectionInstabilities} tells us that we should expect instabilities in the range 
(\ref{case1}) whenever $\gamma = - \alpha$.
As usual, to satisfy NEC one must further impose (\ref{NEC}).

\item
That these spatially modulated instabilities are quite generic, at least in the sense that models that exhibit them 
are easy to construct, should be by now clear.
As one last example to further illustrate this point, consider
\beq
\label{3expV}
V(\phi) = V_0 e^{-\eta\phi} + V_1 e^{\gamma\phi}  + V_2 e^{\delta\phi} \,  .
\eq
Requiring $AdS_2 \times \mathbb{R}^2$ in the IR in this case gives
\beq
\label{3expv2}
v_2 = -L^2 V_2 e^{\delta \phi_0} (\eta+\delta)(\gamma-\delta) + \alpha (\gamma - \eta) - \eta \gamma \, .
\eq
Looking for simplicity again at 
the instability window 
(\ref{case1}), recall that it will be valid provided that (\ref{case1cond}) is obeyed.
For this potential, this boils down to choosing $V_2$ so that 
\beq
V_2 = \frac{\alpha^2 + \alpha(\gamma-\eta)-\gamma\eta }{L^2 e^{\delta\phi_0}(\eta+\delta)(\gamma-\delta)} \, ,
\eq
a condition which should be easy to satisfy -- apart from potential pathologies -- irrespective of the values 
of the remaining Lagrangian parameters.
\item
Finally, we would like to point out that when
the scalar potential is a single exponential, 
\beq
V(\phi) = V_0 e^{-\eta \phi} \,  ,
\eq
the requirement of $AdS_2 \times \mathbb{R}^2$ in the infrared forces $\alpha = \eta$.
We then have 
\beq
\tau_1 = \alpha \, , \quad \quad \quad v_2 = \tau_2 = -\alpha^2 \, ,
\eq
which `trivially' satisfies $\tau_2 + v_2 =- 2 \tau_1^2$ and is therefore an example of \emph{Case 1}. 
However, when $\alpha = \eta$ the solution corresponds to $z = \infty$ and is just $AdS_2 \times \mathbb{R}^2$ 
everywhere, with no intermediate regime of the type we have been describing. 
\end{enumerate}
\vspace{0.3 cm}

\section{Final Remarks}

For $\{z,\theta \}$ scaling solutions with an $AdS_2$ IR completion, the presence of finite-momentum modes which trigger instabilities 
in the near-horizon region can be used to argue that the full geometry should be expected to be unstable to 
spatially modulated phases. 
This logic led us to map the $AdS_2$ instability windows (\ref{window}) and (\ref{Ex1inst}) 
we identified in Section \ref{SectionInstabilities} to conditions 
for the parameters $z$ and $\theta$ characterizing the intermediate scaling regime 
-- in a small-$k$ approximation in (\ref{approx}) and for a simple exact case in (\ref{case1}).
Depending on the detailed structure of the scalar potential,  
there will be regions of phase space in which such instability conditions are satisfied -- in addition, 
models obeying these relations can be engineered in a straightforward manner.

A drawback of this analysis 
is that it doesn't allow one to make `universal' statements about which $\{z,\theta\}$
scaling solutions will be unstable, without fully specifying the scalar potential -- knowledge of the 
infrared behavior of the geometry is clearly not enough to uniquely determine the intermediate scaling region.
Moreover, while the perturbations we used here are more generic than those of \cite{Cremonini:2012ir}, 
we have relied mostly on a small-$k$ expansion, 
and therefore have not identified  all possible sources of instability.

Nonetheless, the analysis of this note reinforces 
the idea that spatially modulated phases seem to arise generically in the deep IR 
of a large class of scaling solutions with hyperscaling violation.
While it would be interesting to have an explicit supergravity realization of the types of flows advocated here, 
there shouldn't be any fundamental obstacle to finding them.
More broadly, it would be useful to have a better understanding of what differentiates between 
the possible IR completions of these classes of scaling solutions, and in particular whether they approach 
$AdS_2 \times \mathbb{R}^2$ or a geometry conformal to it, or even more general classes of anisotropic geometries --
all cases associated with very different physics and transport properties.
Further exploring the rich IR behavior of scaling solutions of this type -- 
and properties of anisotropic and/or inhomogeneous ground states more generally -- will undoubtedly continue 
to bring fresh insights into strongly correlated phases of matter.

\begin{acknowledgments} 
I would like to thank A. Donos, J. Gauntlett and B. Gout\'eraux for valuable discussions.
I am grateful to the Isaac Newton Institute for Mathematical Sciences for hospitality during
the final stages of this project.
This work has been supported by the Cambridge-Mitchell Collaboration in Theoretical Cosmology,
and the Mitchell Family Foundation.
\end{acknowledgments}

\end{document}